\documentclass{article}

\usepackage{amsmath, amsthm, amssymb}
\usepackage{graphicx}
\usepackage{graphics}
\usepackage{subfig}
\usepackage{esint}
\usepackage[a4paper]{geometry}
\begin{document}

\title{Development of a Cox-Thompson inverse scattering method to charged particles}
\author{Tam\'as P\'almai$^1$, Barnab\'as Apagyi$^1$ and Werner Scheid$^2$\\
$^1$ Department of Theoretical Physics\\ Budapest University of
Technology and Economics, H-1111, Budapest, Hungary\\ \ \\
$^2$ Institut f\"ur Theoretische Physik der
Justus-Liebig-Universit\"at, Giessen, Germany\\ \ \\
Electronic mail: palmai@phy.bme.hu, apagyi@phy.bme.hu}

\maketitle

\begin{abstract}

Cox-Thompson fixed-energy quantum inverse scattering method is
developed further to treat long range  Coulomb interaction.
Depending on the reference potentials chosen, two methods have
been formulated which produce inverse potentials with singular or
finite value at the origin. Based on the quality of reproduction
of input experimental phase shifts, it is guessed that the
$p-\alpha$ interaction possesses an interesting repulsive hard
core.\\

\ \\
\noindent PACS: 02.30.Gp, 02.30.Zz, 02.60.Cb,  03.65.Nk, 24.10.-i, 25.40.Cm, 25.40.Ep

\end{abstract}

\section{Introduction} Inverse quantum scattering methods
\cite{Chadan,Geramb,Balaton1996,Balaton2007} represent useful
techniques to assess effective interactions between colliding
composite particles in a model independent way. In nuclear physics
the modified Newton-Sabatier (mNS) method \cite{MS,MMS} has been
used to determine, in general, complex valued optical potentials
describing various systems, e.g., $^{12}$C -- $^{12}$C elastic
scattering \cite{ApagyiJPG91,PRC 1994} or to guess the spin-orbit
potentials arising in the $p$ -- $\alpha$ and $n$ -- $\alpha$
collisions \cite{AlexanderPRC96}. The Newton-Sabatier (NS) method
\cite{RN,Sabatier} has the property that it requires an infinite
set of phase shifts otherwise the first moment of the potential
generated vanishes. This induces an unpleasant oscillation of the
potential at large distances, in addition to the artificial
singularity at the origin which remains to be present in the
potentials generated also by the mNS method.

In the recent decade another procedure, the Cox-Thompson (CT) inverse quantum scattering
method \cite{CT} has been investigated \cite{jpa03,melch,palm1}. This
method has the advantage that it requires a finite set of phase
shifts and possesses a non-zero first momentum of the potential
generated. In addition to this, in general, it also reproduces test potentials better than the mNS method \cite{jpa03}.
The CT method has been applied so far to uncharged
particle scattering in order to reconstruct short ranged
potentials from synthetic phase shifts \cite{jpa03}, and also to
construct $n$ -- $\alpha$ potential from experimental data \cite{melch}.
We shall develop here the method further with the
intention that the CT method can be applied to charged particle
scattering when the potential to be constructed has a long ranged
Coulomb tail. The various extensions will be applied to both
synthetic and experimental phase shifts. The results are obtained by using the nonlinear solver package of MATLAB.

\section{Cox-Thompson (CT) method}

The original CT method belongs to the fixed-energy procedures when a finite set of $N$
phase shifts $\{\delta_l\}_{l\in S}$ is converted to the spherical
potential $V(r)$ appearing in the $l-$th partial wave radial
Schr\"odinger equation
\begin{equation}\label{schr}
\left(-\frac{\hbar^2}{2m}\frac{ d^2}{ d
r^2}+\frac{\hbar^2}{2m}\frac{l(l+1)}{r^2}+V(r)-\frac{\hbar^2}{2m}k^2\right){\Psi}_l(k;r)=0
\end{equation}
with boundary conditions $\Psi_l(k;0)\propto r^{l+1}$ and
$\Psi_l(k;r\to\infty)\propto \sin(kr-l{\pi \over 2}+\delta_l).$

We shall also refer to another form of the Schr\"odinger equation
\begin{equation}\label{dimsch}
x^2\left(\frac{ d^2}{ d x^2}+1-q(x)\right)\psi_l(x)=l(l+1)\psi_l(x),
\end{equation}
where dimensionless quantities have been introduced
for the distance  $x=kr$, for the potential
$q(x)=E^{-1}V(xk^{-1})$ with $E=\hbar^2k^2/(2m)$, and  the dimensionless wave function
 $\psi_l(x)$  is related to the wave function  $\Psi_l(k;r)$ through the asymptotic normalization constant
$\Psi_l(k;r)=C_l \psi_l(kr)$.

 The conversion from phase shifts $\{\delta_l\}$ to
potential $V(r)$ can be carried out in two steps
\cite{jpa03,melch}. First, we solve either of the following two
systems of highly non-linear equations
\begin{equation}\label{general}
 e^{2 i\delta_l}=\frac{1+ i\mathcal{K}_l^+}{1- i\mathcal{K}_l^-}\qquad
{\rm{or}} \qquad\
\tan(\delta_l)=\frac{\mathcal{K}_l^++\mathcal{K}_l^-}{2+ i(\mathcal{K}_l^+-\mathcal{K}_l^-)},
\qquad l\in S,
\end{equation}
with
\begin{equation}
\mathcal{K}_l^{\pm}=\sum_{L\in T, l'\in
S}[M_{\sin}]_{lL}[M_{\cos}^{-1}]_{Ll'} e^{\pm  i(l-l')\pi/2}, \qquad
l\in S.
\end{equation}
and
\begin{equation}\label{Msincos}
\left\{\begin{array}{ll}
M_{\sin}\\
M_{\cos}
\end{array}\right\}_{lL} = \frac{1}{L(L+1)-l(l+1)}\left\{ \begin{array}{ll}
\sin\left((l-L)\frac{\pi}{2}\right)\\
\cos\left((l-L)\frac{\pi}{2}\right)
\end{array} \right\},\qquad l\in S,\,L\in T
\end{equation}
for the set $T$ of the shifted angular momenta $\{L\}$ where the
relations $S\cap T=\emptyset$ and $|T|=|S|=N$ hold. Next, one
calculates the potential  as
\begin{equation}
V(r)=Eq(kr),
\end{equation}
where
 \begin{equation}
q(x)=-\frac{2}{x}\frac{ d}{ d x}\left(\frac{K(x,x)}{x}\right)
\end{equation}
with
\begin{equation}\label{CTanK}
K(x,y)=\sum_{L \in T}A_L(x)u_L(y),
\end{equation}
where the coefficient functions $A_L(x)$ are calculated by solving
the system of linear equations
\begin{equation}\label{AL}
\sum_{L\in
T}A_L(x)\frac{W[u_L(x),v_l(x)]}{l(l+1)-L(L+1)}=v_l(x),\qquad l\in
S
\end{equation}
with
\begin{equation}
u_n(x)=\sqrt{\frac{\pi x}{2}}J_{n+\frac{1}{2}}(x),\qquad
v_n(x)=\sqrt{\frac{\pi x}{2}}Y_{n+\frac{1}{2}}(x),\qquad n\in
S\cup T
\end{equation}
being the regular and irregular solutions of the dimensionless free radial Schr\"odinger equation for the $n$th partial wave (Riccati-Bessel functions \cite{abramo}). Note that the set $T$ of shifted angular momenta $L$ contains real or (in case of complex phase shifts) even complex numbers.

Using the behaviour of these functions at the origin and infinity,
one easily derives the following  properties of the Cox-Thompson
potential. At the origin  we get \cite{jpa03}, in general, a
constant
\begin{equation}
q(0)=Q-2(1-Q)\sum_{L\in T}\sum_{l\in S}\frac{G^{-1}_{lL}}{2l-1}.
\end{equation}
with the matrix $G_{Ll}=(L-l)^{-1}$ and the constant quantity
$Q=\sum_{L\in T}\sum_{l\in S}\frac{G^{-1}_{lL}}{L+\frac{3}{2}}$.

At the infinity the CT potential in general falls off like an inverse power
of two
\begin{equation}
 q(x\to\infty)=-\frac{1}{{x}^2}(\alpha_T\cos(2x)-\beta_T\sin(2x))
+\frac{2}{{x}^3}(\alpha_T\sin(2x)+
\beta_T\cos(2x)+\gamma_T)
\end{equation}
where the coefficients depending on the set $T$ are defined by the
following relations
\begin{eqnarray}
\alpha_T=\frac{1}{2}\sum_{L\in T}\left(a_L\cos L\frac{\pi}{2}-b_L\sin L\frac{\pi}{2}\right),\\
\beta_T=-\frac{1}{2}\sum_{L\in T}\left(b_L\cos L\frac{\pi}{2}+a_L\sin L\frac{\pi}{2}\right),\\
\gamma_T=\frac{1}{2}\sum_{L\in T}\left(b_L\cos
L\frac{\pi}{2}-a_L\sin L\frac{\pi}{2}\right),
\end{eqnarray}
with the coefficients $a_L, b_L$ defined via the asymptotic
expansion functions $A_L^{\rm{a}}(x)\equiv A_L(x\to\infty)$ which
admits a periodic expansion \cite{palm1} as
\begin{equation}\label{difAsol}
A_L^{\rm{a}}(x)=a_L\cos(x)+b_L\sin(x)
\end{equation}
because of the form of the asymptotic version of equation (\ref{AL})
\begin{equation}\label{ALa}
\sum_{L\in
T}A_L^{\rm{a}}(x)\frac{\cos((l-L)\frac{\pi}{2})}{l(l+1)-L(L+1)}=
-\cos(x-l\frac{\pi}{2}),\qquad l\in S.
\end{equation}

\section{Extensions of CT method to long ranged potential}
In nuclear physics one frequently encounters  charged particle scattering.
The appearance of the long ranged Coulomb potential requires an
appropriate extension of the CT method which will be carried out
in various ways in this section. All the different extensions rely
upon the two-potential formalism \cite{Joahain}.

Consider  a spherical potential $V(r)$ which can be written as a
sum of a short-ranged interior (or nuclear) part $\hat V$ and a long-ranged
exterior (asymptotic or reference) part $V^{(0)}(r)$, i.e.,
\begin{equation}\label{splitpot}
V(r)= \hat V(r)+ V^{(0)}(r).
\end{equation}
 Then the phase shift itself is also split
into two parts
\begin{equation}
\delta_l=\hat\delta_l+\delta_l^{(0)}
\end{equation}
where $\delta_l^{(0)}$ is the phase shift due to $V^{(0)}$ and $\hat\delta_l$ corresponds to the phase shifts caused by
$\hat V$ (in the presence of $V^{(0)}$). If, for instance, $\hat{V}(r)\equiv0$, $\hat{\delta}_l=0$ $\forall l$.

If the short ranged part of the potential is zero beyond a finite
distance $r_{\rm a}$,
\begin{equation}\label{hatV}
\hat V(r)=0, \quad r\geq r_{\rm a}
\end{equation}
then, accordingly, the radial  wave function can be written  in
this region as
\begin{equation}\label{psiref}
\Psi_l(k;r\geq r_{\rm a})=C_l\left( F_l^{(0)}(kr) +  \tan \hat \delta_l
G_l^{(0)}(kr) \right).
\end{equation}
In the latter equation the functions  $F_l^{(0)}$ and $G_l^{(0)}$
mean, respectively, the regular and irregular solutions of the dimensionless
radial Schr\"odinger equation (\ref{dimsch}) with potential $q(x)=E^{-1}V^{(0)}(xk^{-1})$.

\subsection{Phase transformation method (PCT)}

We may design a modified potential $\tilde V(r) $ which is
identical with the original one (up to a constant energy shift) within
the interior region $0<r<r_{\rm a}$ and zero outside
\begin{equation}\label{splitpot-1}
\tilde V(r)=\left\{\begin{array}{ll}
 V(r)-V(r_{\rm a}), & r\leq r_{\rm a},\\
 0, & r\geq r_{\rm a}.
\end{array} \right.
\end{equation}
Thus if equation (\ref{hatV}) applies the $\hat{V}$ interior potential can be deduced from $\tilde{V}$.
The radial scattering wave function of this problem at the shifted
energy $E_{\rm B}=E-V(r_{\rm a})\equiv \hbar^2k_{\rm B}^2/2m$ is identical with the
original one in the inside region
\begin{equation}
\tilde\Psi_l(k_{\rm B};r)=\Psi_l(k;r),\quad r\leq r_{\rm a}
\end{equation}
but differs in the outside region
\begin{equation}
\tilde\Psi_l(k_{\rm B};r)=\tilde C_l\left(
u_l(k_{\rm B}r)-\tan{\tilde\delta}_lv_l(k_{\rm B}r) \right),\quad r \geq r_{\rm a}.
\end{equation}
The equality of the logarithmic derivatives at $r=r_{\rm a}$
\begin{equation}
\left.\frac{ d}{ d
r}\log\left(\frac{1}{r}\tilde\Psi_l(k_{\rm B};r)\right)\right|_{r=r_{\rm a}}=\left.\frac{ d}{ d
r}\log\left(\frac{1}{r}\Psi_l(k;r)\right)\right|_{r=r_{\rm a}}
\end{equation}
with $\Psi_l(k;r)$ from (\ref{psiref})
 gives the transformed phase shifts $\tilde\delta_l(k_{\rm B})$
which can be used to (re)construct the short ranged   potential
$\tilde V(r)$ (\ref{splitpot-1}). To get the whole original
potential one simply adds the reference potential
\begin{equation}\label{refsplitpot-1}
\tilde V^{(0)}(r)=\left\{\begin{array}{ll}
 V(r_{\rm a}), & r\leq r_{\rm a},\\
  V^{(0)}(r), & r\geq r_{\rm a},
\end{array} \right.
\end{equation}
if it is known.

The phase shift transformation method enables the use of the CT
method as formulated in the preceding section. It requires that
the modified potential $\tilde V(r)$ be constructed from the set $\{\tilde{\delta}_l\}$ by the actual
inverse method as accurately as possible. But no exact fixed energy
inverse scattering method has yet been formulated which is able to
produce a short range potential that is exactly zero beyond a finite
radius, if a finite set of phase shifts is employed. Therefore, one
always introduces a small error into the potential
(re)construction when applying the phase shift transformation
procedure.

The phase shift transformation method has been introduced by May,
M\"unchow and Scheid \cite{MMS} and is widely applied in case of
the modified Newton Sabatier method. It is applied also to the CT
inverse procedure to treat Coulomb scattering
\cite{Melchertthesis} and we shall call this combined method as the
PCT method.

\subsection{Generalized CT scheme}

Based on the two potential formalism one can derive a generalized
 CT (gCT) scheme which employs the given  nuclear phase shifts $\{
\hat\delta_l\}$ for constructing the short-ranged (nuclear)
potential $\hat V(r)$ but avoids use of a matching radius $r_{\rm a}$. The
derivation starts with the  ansatz for the input symmetrical kernel of the Gel'fand-Levitan-type integral equation
\begin{equation}
g(x,x')=\sum_{l\in S}\gamma_l F_l^{(0)}(x_<)G_l^{(0)}(x_>),
\end{equation}
where $x_<(x_>)$ denotes the lesser (greater) of $x,x'$. Then, by
proceeding through the usual steps \cite{jpa03,melch}, one arrives
at a system of nonlinear equations identical in structure to (\ref{general})
\begin{equation}\label{general-c}
 e^{2 i\hat\delta_l}=\frac{1+ i\mathcal{K}_l^+}{1- i\mathcal{K}_l^-}\qquad
{\rm{or}} \qquad\
\tan(\hat\delta_l)=\frac{\mathcal{K}_l^++\mathcal{K}_l^-}{2+ i(\mathcal{K}_l^+-\mathcal{K}_l^-)},
\qquad l\in S,
\end{equation}
with
\begin{equation}
\mathcal{K}_l^{\pm}=\sum_{L\in T, l'\in
S}[M_{\sin}]_{lL}[M_{\cos}^{-1}]_{Ll'} e^{\pm
 i\left[(l-l')\pi/2+\delta_{l'}^{(0)}-\delta_l^{(0)}\right]}, \qquad l\in S,
\end{equation}
and
\begin{equation}\label{Msincos-c}
\left\{\begin{array}{ll}
M_{\sin}\\
M_{\cos}
\end{array}\right\}_{lL} = \frac{1}{L(L+1)-l(l+1)}\left\{ \begin{array}{ll}
\sin\left((l-L)\frac{\pi}{2}+\delta_L^{(0)}-\delta_l^{(0)}\right)\\
\cos\left((l-L)\frac{\pi}{2}+\delta_L^{(0)}-\delta_l^{(0)}\right)
\end{array} \right\},\qquad l\in S,\,L\in T
\end{equation}
for the set $T$ of the shifted angular momenta $L$ where the
relations $S\cap T=\emptyset$ and $|T|=|S|$ hold. From the set $T$
one calculates the nuclear potential as
\begin{equation}
\hat V(r)=Eq(kr),
\end{equation}
where
 \begin{equation}
q(x)=-\frac{2}{x}\frac{ d}{ d x}\left(\frac{K(x,x)}{x}\right)
\end{equation}
with
\begin{equation}\label{CTanK-c}
K(x,y)=\sum_{L \in T}A_L(x)F^{(0)}_L(y).
\end{equation}
The coefficient function $A_L(x)$ is calculated by solving the
system of linear equations
\begin{equation}\label{AL-c}
\sum_{L\in
T}A_L(x)\frac{W[F^{(0)}_L(x),G^{(0)}_l(x)]}{l(l+1)-L(L+1)}=G^{(0)}_l(x),\qquad
l\in S,
\end{equation}

From this gCT scheme several developments are
possible as the actual form of the reference potential has not yet been specified.

\subsection{Coulomb reference potential method (CCT)}
If one sets the reference potential to be the bare Coulomb
potential
\begin{equation}
V^{(0)}(r)=E{2\eta\over kr}\equiv
\frac{1}{4\pi\varepsilon_0}\frac{Z_1Z_2e^2}{r},
\end{equation}
with $\eta $ being the Sommerfeld parameter then one arrives at
the Coulomb CT (CCT) method. In this case the regular and
irregular reference functions are the regular and irregular
Coulomb functions, the reference phase shift becomes the Coulomb
phase
\begin{equation}\label{Cphase}
\sigma_l=\frac{1}{2 i}\ln\left[\frac{\Gamma(l+1+ i\eta)}{\Gamma(l+1- i\eta)}\right],\qquad
l\in S.
\end{equation}

Here it should be mentioned that in the course of application of CCT  method
it may become necessary to know the regular and irregular Coulomb functions and Coulomb phases for
complex orders. For example, in case of non-elastic scattering the phases are complex and therefore, as noted before,
the $L$ numbers are also complex valued. The Coulomb functions are well-defined for complex orders
and similarly to the real order case they can be given as power series (for details see e.g. \cite{coulFG,coulFG2} and Appendix).

It is interesting that contrary to the fact that
the asymptotic form of the reference functions contains the well
known logarithmic term $-\eta\ln 2kr$ in the argument, it does not
appear in the gCT formulas because of cancelation.

Using the known power series of the Coulomb functions it can be shown that the
CCT method gives a potential which is proportional to the Coulomb potential near the origin. For one term,
i.e.  $|T|=|S|=1$,
 we get
\begin{equation}
V(r\approx 0)=
E\left[\frac{L(1+l)}{l(1+L)}\right]\frac{2\eta}{kr}+O(1).
\end{equation}
For  large $r$ we get
\begin{equation}
 V(r\to\infty)=
E{2\eta\over kr}
-\frac{2E}{{(kr)}^2}\sum_{L\in T}\sum_{l \in
S}[M_{\cos}^{-1}]_{Ll}\cos(\Theta_L(kr)+\Theta_l(kr))+O\left(\frac{1}{{(kr)}^3}\right),
\end{equation}
with $\Theta_L(x)=x-\eta\ln2x-L\frac{\pi}{2}+\sigma_L$.

We see that the CCT method generates an inverse potential that
 gives a Coulomb-like singularity at the origin and produces a damped oscillation around the  Coulomb
tail at large distances. It is free of the matching parameter
$r_{\rm a}$ and requires just the nuclear phase shifts $\hat\delta_l$
which are derived by the usual phase shift analysis procedures.

\subsection{Modified Coulomb reference potential method (MCT)}

In order to obtain an inverse potential that is finite at the
origin, instead of being singular there,  we can modify the
Coulomb reference potential accordingly. This reference potential
is constant in the interior domain and purely Coulombic outside.
This modified Coulomb potential is the same as that employed by the
phase transformation method for Coulomb asymptotics and reads as
\begin{equation}
V^{(0)}(r)=\left\{\begin{array}{ll}E\frac{2\eta}{kr_{\rm a}},&\textrm{$r\leq
r_{\rm a}$,}\\ E\frac{2\eta}{kr},&\textrm{$r\geq
r_{\rm a}$}\end{array}\right..
\end{equation}
To this reference potential there belong the following regular and
irregular  radial wave function
\begin{eqnarray}
F_l^{(0)}(kr)=\left\{\begin{array}{ll}u_l\left(\sqrt{1-\frac{2\eta}{kr_{\rm a}}}\cdot
kr\right),
&\textrm{$r< r_{\rm a}$,}\\ \alpha_{F_l^{(0)}}F_l(kr)+\beta_{F_l^{(0)}}G_l(kr),&\textrm{$r> r_{\rm a}$,}\end{array}\right.\\
G_l^{(0)}(kr)=\left\{\begin{array}{ll}v_l\left(\sqrt{1-\frac{2\eta}{kr_{\rm a}}}\cdot
kr\right),&\textrm{$r< r_{\rm a}$,}\\
\alpha_{G_l^{(0)}}F_l(kr)+\beta_{G_l^{(0)}}G_l(kr),&\textrm{$r>
r_{\rm a}$,}\end{array}\right.
\end{eqnarray}
and reference phase shift
\begin{equation}
\delta^{(0)}_l=\sigma_l+\arctan\left(\frac{\beta_{F_l^{(0)}}}{\alpha_{F_l^{(0)}}}\right).
\end{equation}
The  coefficients $\alpha_{F_l^{(0)}},
\beta_{F_l^{(0)}},\alpha_{G_l^{(0)}},\beta_{G_l^{(0)}}$ can be calculated from the
equality of the  inner and outer wave functions and their
derivatives at the matching radius $r_{\rm a}$. For example, the two
coefficients necessary for calculating the reference phase shift
are
 \begin{equation}
 \alpha_{F_l^{(0)}}=\frac{\frac{u_l\left(\sqrt{1-\frac{2\eta}{x_{\rm a}}}\cdot
x_{\rm a}\right)}{G_l(x_{\rm a})}-\frac{\sqrt{1-\frac{2\eta}{x_{\rm a}}}u'_l\left(\sqrt{1-\frac{2\eta}{x_{\rm a}}}\cdot
x_{\rm a}\right)}{G'_l(x_{\rm a})}}{\frac{F_l(x_{\rm a})}{G_l(x_{\rm a})}-\frac{F'_l(x_{\rm a})}{G'_l(x_{\rm a})}},\qquad
\beta_{F_l^{(0)}}=\frac{\frac{u_l\left(\sqrt{1-\frac{2\eta}{x_{\rm a}}}\cdot
x_{\rm a}\right)}{F_l(x_{\rm a})}-\frac{\sqrt{1-\frac{2\eta}{x_{\rm a}}}u'_l\left(\sqrt{1-\frac{2\eta}{x_{\rm a}}}\cdot
x_{\rm a}\right)}{F'_l(x_{\rm a})}}{\frac{G_l(x_{\rm a})}{F_l(x_{\rm a})}-\frac{G'_l(x_{\rm a})}{F'_l(x_{\rm a})}},
\end{equation}
with $x_{\rm a}=kr_{\rm a}$ and prime denotes derivation with respect to argument.

The total phase shifts can be written in two different ways
\begin{equation}
\delta_l=\hat\delta_l+\sigma_l=\delta_l^{\rm MCT}+\delta_l^{(0)}
\end{equation}
where $\hat\delta_l$ means the nuclear phase shifts given as data
and $\delta_l^{\rm MCT}$ is to be used to perform the CT inverse
calculation outlined above.

The potential obtained by the MCT method has a finite value at the origin.  Because this method employs a similar reference potential as the PCT method, the results provided by the two methods should also be very similar, although quite different functions are used in the calculations. The advantage of the MCT over the PCT lies in that it does not involve the small error in the phase shift reproduction inherent to the PCT.

\section{Applications to (re)construct effective potentials}

\subsection{Synthetic phase shifts}

To illustrate the general applicability of the long-range CT inversion procedures we shall first use them to reconstruct model potentials. We model the $\alpha$ -- $\alpha$ scattering with two slightly different potentials: the first is finite
at the origin
and the second is singular  (describing a possible non-locality).

The first model potential is given by
\begin{equation}\label{43}
V(r)=U(r)+V_{\rm C}(r),
\end{equation}
where the nuclear interaction is described by a Woods-Saxon form
\begin{equation}
U(r)=U_0\left(1+ d^{\frac{r-R}{a}}\right)^{-1},
\end{equation}
and the Coulomb interaction is represented by the potential of a homogeneous charged sphere as
\begin{equation}
V_{\rm C}(r)=\left\{\begin{array}{ll}\frac{Z_1Z_2e^2}{2R_{\rm C}}\left(3-\frac{r^2}{R_{\rm C}^2}\right),&r\leq
R_{\rm C},\\ \frac{Z_1Z_2e^2}{r},&r>R_{\rm C}.\end{array}\right.
\end{equation}

For the various parameters we choose the following values:
$Z_1=2$, $Z_2=2$, $A_1=4$, $A_2=4$, $U_0=-20$ MeV, $R=R_{\rm C}=2.0636$
fm, $a=0.25$ fm.

\begin{figure}[ht]
  \begin{minipage}[b]{0.5\linewidth}
     \centering
     \subfloat[][$E_{\rm c.m.}=25$ MeV]{\includegraphics[width=7cm]{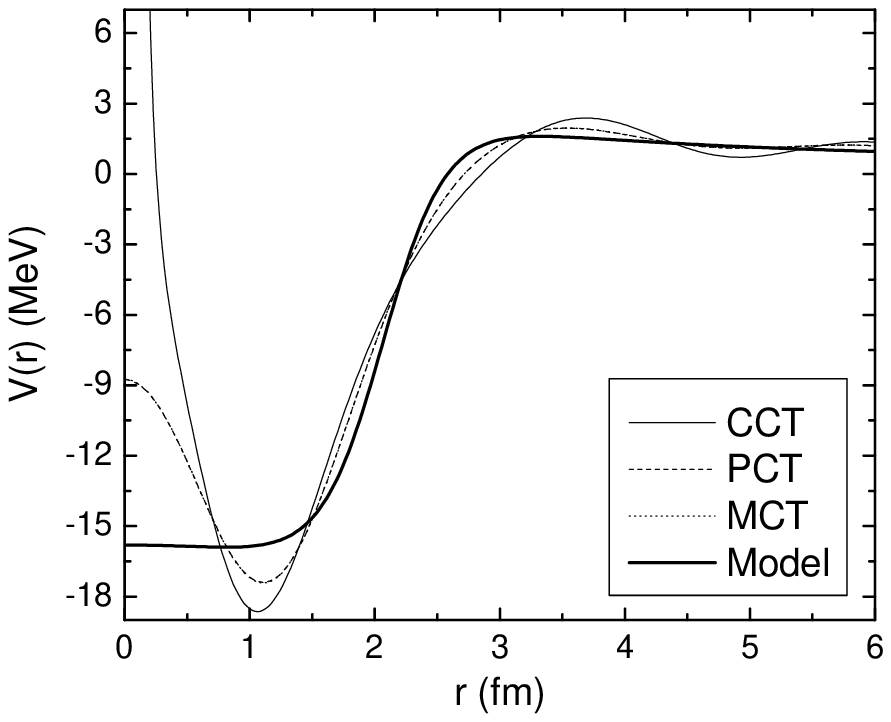}}\\
  \end{minipage}
  \hspace{0.1cm}
  \begin{minipage}[b]{0.5\linewidth}
     \centering
     \subfloat[][$E_{\rm c.m.}=35$ MeV]{\includegraphics[width=7cm]{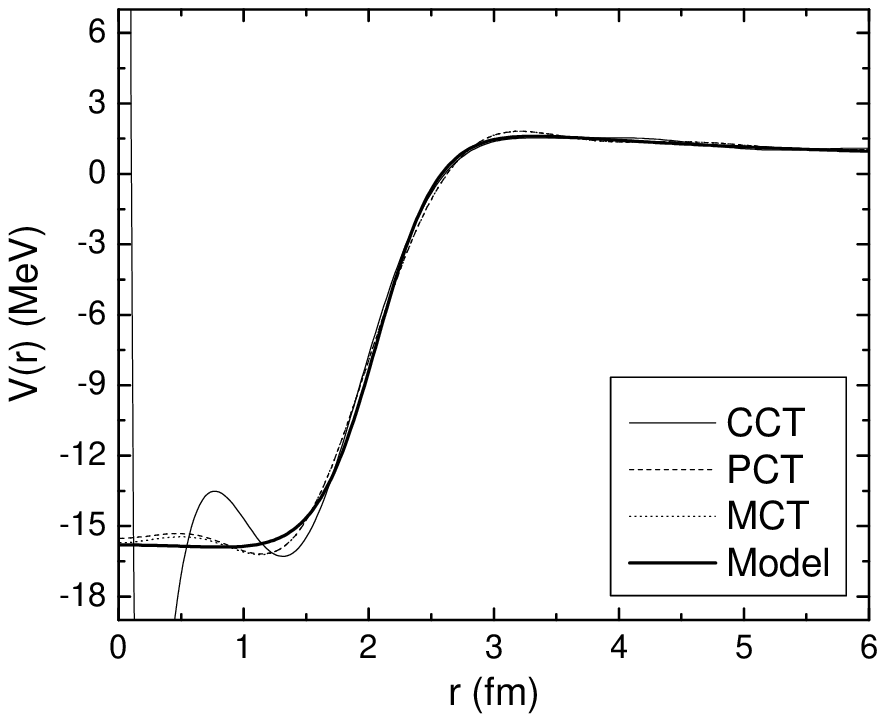}}\\
   \end{minipage}
   \caption{Model with potential (\ref{43}) and inverse potentials
    yielded by the CCT, MCT and PCT methods (labeled accordingly) at $E_{\rm c.m.}=25$ and 35 MeV center of mass energies. The model potential is non-singular at the origin.}
\label{fig0}
\end{figure}

We have calculated the nuclear phase shifts $\hat\delta_l$ at
energies $E=25$ and $35$ MeV (see tables \ref{tab0} and
\ref{tab01}).  These phase shifts are then used as input data for
the various CT calculations. The results for the $L$-values obtained using the nonlinear solver of MATLAB are shown in tables
\ref{tab0} and \ref{tab01}. The corresponding potentials are displayed in figure \ref{fig0}.

\begin{table}[h]
\caption{\label{tab0} Model data ($\hat{\delta}_l^{\rm orig}$
phase shifts),  inversion results ($L$ shifted angular momenta and
$\Delta\hat{\delta}_l$ differences between the model phase shifts
and the ones given by the various methods$\ddag$) of the $\alpha$
-- $\alpha$ scattering at $E_{{\rm c.m.}}=25$ MeV. The matching
parameter used in the PCT and MCT procedures was set to $r_{\rm a}=10$
fm.}

\begin{center}
\begin{tabular}{cccccccc}
\hline
$l$ & $L^{{\rm CCT}}$ & $L^{{\rm MCT}}$ & $L^{{\rm PCT}}$
& $\hat{\delta}_l^{{\rm orig}}$ & $\Delta\hat{\delta}_l^{{\rm CCT}}$&$\Delta\hat{\delta}_l^{{\rm MCT}}$&$\Delta\hat{\delta}_l^{{\rm PCT}}$\\
\hline
$0$  &$-1.5622$  &$-0.9225$ & $-0.9219$     &$1.2989$&  $0.0085$&$0.0053$&$0.0056$\\
$1$ &$\phantom{-}$$0.5286$& $\phantom{-}$$0.5841$ &$\phantom{-}$$0.5843$     &$1.1445$&  $0.0011$&$0.0009$&$0.0001$\\
$2$ &$\phantom{-}$$1.6462$& $\phantom{-}$$1.7503$ &$\phantom{-}$$1.7502$     &$0.8307$&  $0.0116$&$0.0063$&$0.0062$\\
$3$ &$\phantom{-}$$2.9468$& $\phantom{-}$$3.0281$ &$\phantom{-}$$3.0281$     &$0.2300$&  $0.0131$&$0.0025$&$0.0024$\\
$4$ &$\phantom{-}$$4.0456$& $\phantom{-}$$4.1114$ &$\phantom{-}$$4.1115$     &$0.0399$&  $0.0176$&$0.0053$&$0.0054$\\
$5$ &$\phantom{-}$$5.0272$& $\phantom{-}$$5.0983$ &$\phantom{-}$$5.0982$     &$0.0062$&  $0.0080$&$0.0019$&$0.0018$\\
$6$ &$\phantom{-}$$6.0308$& $\phantom{-}$$6.0874$ &$\phantom{-}$$6.0875$     &$0.0009$&  $0.0073$&$0.0004$&$0.0005$\\
$7$ &$\phantom{-}$$7.0169$& $\phantom{-}$$7.0684$ &$\phantom{-}$$7.0684$     &$0.0001$&  $0.0047$&$0.0039$&$0.0039$\\
$8$ &$\phantom{-}$$8.0177$& $\phantom{-}$$8.0557$ &$\phantom{-}$$8.0557$     &$0.0000$&  $0.0001$&$0.0066$&$0.0066$\\
$9$ &$\phantom{-}$$9.0108$& $\phantom{-}$$9.0428$ &$\phantom{-}$$9.0427$     &$0.0000$&  $0.0001$&$0.0108$&$0.0108$\\
\hline
\end{tabular}
\end{center}

$\ddag$ Note that for the sake of comparison the phases given by the inverse potentials were calculated
by cutting-off the non-physical oscillations beyond
the matching radius $r_{\rm a}$ used in the PCT procedure. Without the cut-off the MCT and CCT potentials reproduce the phase shifts within an error
of the numerical precision.

\end{table}

\begin{table}[h]
\caption{\label{tab01} Model data and inversion results of the $\alpha$ -- $\alpha$ scattering at $E_{{\rm c.m.}}=35$ MeV. The matching parameter used in the PCT and MCT procedures was set to $r_{\rm a}=10$ fm.}
\begin{center}
\begin{tabular}{cccccccc}
\hline
$l$ & $L^{{\rm CCT}}$ & $L^{{\rm MCT}}$ & $L^{{\rm PCT}}$ & $\hat{\delta}_l^{{\rm orig}}$ & $\Delta\hat{\delta}_l^{{\rm CCT}}$&$\Delta\hat{\delta}_l^{{\rm MCT}}$&$\Delta\hat{\delta}_l^{{\rm PCT}}$\\
\hline
$0$   &\phantom{0}$-1.4866$  &\phantom{0}$-0.8389$ & \phantom{0}$-0.8361$     &$1.1880$&  $0.0032$&$0.0002$&$0.0033$\\
$1$  &\phantom{0}\phantom{-}$0.4954$& \phantom{0}\phantom{-}$0.6158$ &\phantom{0}\phantom{-}$0.6161$     &$0.9864$&  $0.0023$&$0.0011$&$0.0001$\\
$2$  &\phantom{0}\phantom{-}$1.6375$& \phantom{0}\phantom{-}$1.7077$ &\phantom{0}\phantom{-}$1.7076$     &$0.8542$&  $0.0032$&$0.0027$&$0.0027$\\
$3$  &\phantom{0}\phantom{-}$2.8420$& \phantom{0}\phantom{-}$2.9144$ &\phantom{0}\phantom{-}$2.9144$     &$0.4246$&  $0.0069$&$0.0031$&$0.0033$\\
$4$  &\phantom{0}\phantom{-}$4.0002$& \phantom{0}\phantom{-}$4.0677$ &\phantom{0}\phantom{-}$4.0677$     &$0.1111$&  $0.0030$&$0.0017$&$0.0016$\\
$5$  &\phantom{0}\phantom{-}$5.0256$& \phantom{0}\phantom{-}$5.0873$ &\phantom{0}\phantom{-}$5.0873$     &$0.0231$&  $0.0079$&$0.0043$&$0.0044$\\
$6$  &\phantom{0}\phantom{-}$6.0277$& \phantom{0}\phantom{-}$6.0873$ &\phantom{0}\phantom{-}$6.0873$     &$0.0043$&  $0.0001$&$0.0002$&$0.0002$\\
$7$  &\phantom{0}\phantom{-}$7.0199$& \phantom{0}\phantom{-}$7.0713$ &\phantom{0}\phantom{-}$7.0713$     &$0.0008$&  $0.0032$&$0.0019$&$0.0019$\\
$8$  &\phantom{0}\phantom{-}$8.0176$& \phantom{0}\phantom{-}$8.0627$ &\phantom{0}\phantom{-}$8.0627$     &$0.0001$&  $0.0047$&$0.0029$&$0.0030$\\
$9$  &\phantom{0}\phantom{-}$9.0124$& \phantom{0}\phantom{-}$9.0492$ &\phantom{0}\phantom{-}$9.0492$     &$0.0000$&  $0.0011$&$0.0009$&$0.0009$\\
$10$ &\phantom{-}$10.0115$&\phantom{-}$10.0421$&\phantom{-}$10.0420$     &$0.0000$&  $0.0034$&$0.0030$&$0.0031$\\
$11$ &\phantom{-}$11.0083$&\phantom{-}$11.0315$&\phantom{-}$11.0316$     &$0.0000$&  $0.0051$&$0.0044$&$0.0044$\\
$12$ &\phantom{-}$12.0082$&\phantom{-}$12.0267$&\phantom{-}$12.0267$     &$0.0000$&  $0.0007$&$0.0067$&$0.0067$\\
\hline
\end{tabular}
\end{center}
\end{table}

 One can observe that as expected the
 PCT and MCT procedures give almost the same results and
 the CCT inverse potential is divergent at the origin.
 By increasing the scattering energy more  phase shifts become
 available for the inversion (see tables \ref{tab0} and \ref{tab01})
 and the potential reproduction becomes better.

\begin{figure}[ht]
\begin{center}
\rotatebox{0}{
 \includegraphics[width=11cm]{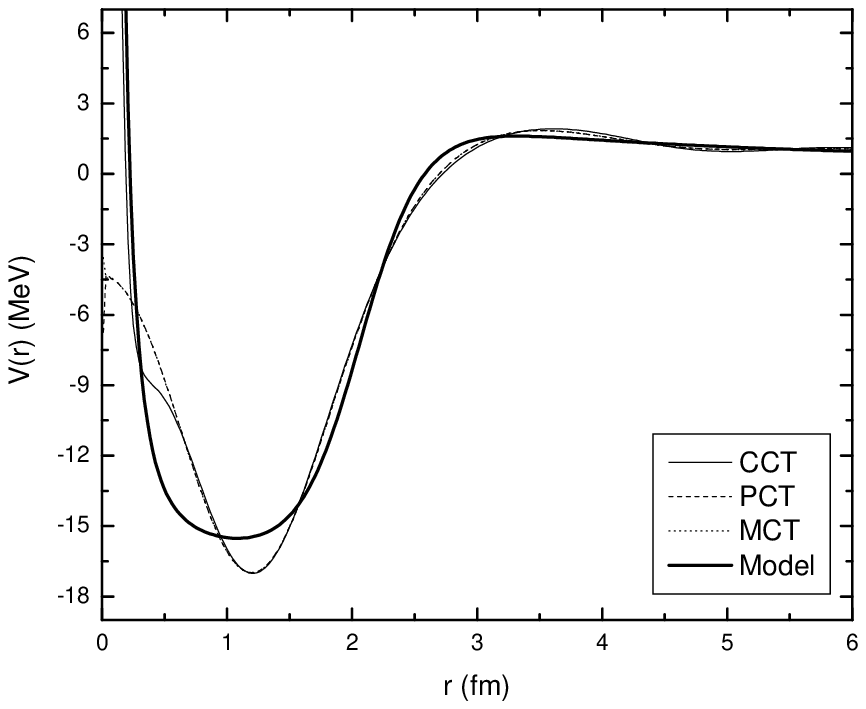}
}
   \caption{Model with  potential (\ref{47}) and inverse potentials yielded by the CCT, MCT and PCT methods (labeled accordingly)
   at $E_{\rm c.m.}=25$ MeV center of mass energy. The model potential is singular at the origin.}
\label{figs}
\end{center}
\end{figure}

The second model is obtained by adding the singular potential term
\begin{equation}
V_{\rm sing}(r)=\frac{ d^{-r}}{r^2}
\end{equation}
to the previous model, i.e.,
\begin{equation}\label{47}
V(r)=U(r)+V_{\rm C}(r)+V_{\rm sing}(r).
\end{equation}
 The reconstruction of phases is listed in table \ref{tab0s}
 and the potentials are shown in figure \ref{figs}.
 We see that the CCT potential follows nicely the model potential in the singular domain
 near the origin while the PCT and MCT methods are unable to reproduce the singularity
 although their phase shift reconstruction is good.

\begin{table}[h]
\caption{\label{tab0s} Model data and inversion results for the divergent model potential at $E_{{\rm c.m.}}=25$ MeV. The matching parameter used in the PCT and MCT procedures was set to $r_{\rm a}=10$ fm.} 
\begin{center}
\begin{tabular}{cccccccc}
\hline
$l$ & $L^{{\rm CCT}}$ & $L^{{\rm MCT}}$ & $L^{{\rm PCT}}$ & $\hat{\delta}_l^{{\rm orig}}$ & $\Delta\hat{\delta}_l^{{\rm CCT}}$&$\Delta\hat{\delta}_l^{{\rm MCT}}$&$\Delta\hat{\delta}_l^{{\rm PCT}}$\\
\hline
$0$   &\phantom{0}$-1.5257$  &\phantom{0}$-0.8526$ & \phantom{0}$-0.8522$     &$1.2121$&  $0.0046$&$0.0038$&$0.0041$\\
$1$  &\phantom{0}\phantom{-}$0.5116$& \phantom{0}\phantom{-}$0.5862$ &\phantom{0}\phantom{-}$0.5862$     &$1.1314$&  $0.0011$&$0.0003$&$0.0007$\\
$2$  &\phantom{0}\phantom{-}$1.6606$& \phantom{0}\phantom{-}$1.7506$ &\phantom{0}\phantom{-}$1.7505$     &$0.8245$&  $0.0051$&$0.0043$&$0.0042$\\
$3$  &\phantom{0}\phantom{-}$2.9505$& \phantom{0}\phantom{-}$3.0290$ &\phantom{0}\phantom{-}$3.0290$     &$0.2284$&  $0.0061$&$0.0028$&$0.0027$\\
$4$  &\phantom{0}\phantom{-}$4.0401$& \phantom{0}\phantom{-}$4.1108$ &\phantom{0}\phantom{-}$4.1109$     &$0.0395$&  $0.0094$&$0.0060$&$0.0060$\\
$5$  &\phantom{0}\phantom{-}$5.0293$& \phantom{0}\phantom{-}$5.0978$ &\phantom{0}\phantom{-}$5.0977$     &$0.0060$&  $0.0036$&$0.0013$&$0.0013$\\
$6$  &\phantom{0}\phantom{-}$6.0283$& \phantom{0}\phantom{-}$6.0870$ &\phantom{0}\phantom{-}$6.0871$     &$0.0008$&  $0.0040$&$0.0031$&$0.0031$\\
$7$  &\phantom{0}\phantom{-}$7.0179$& \phantom{0}\phantom{-}$7.0677$ &\phantom{0}\phantom{-}$7.0677$     &$0.0001$&  $0.0030$&$0.0023$&$0.0023$\\
$8$  &\phantom{0}\phantom{-}$8.0163$& \phantom{0}\phantom{-}$8.0553$ &\phantom{0}\phantom{-}$8.0553$     &$0.0000$&  $0.0001$&$0.0008$&$0.0008$\\
$9$  &\phantom{0}\phantom{-}$9.0113$& \phantom{0}\phantom{-}$9.0413$ &\phantom{0}\phantom{-}$9.0413$     &$0.0000$&  $0.0020$&$0.0020$&$0.0020$\\
$10$ &\phantom{-}$10.0105$&\phantom{-}$10.0325$&\phantom{-}$10.0325$          &$0.0000$&  $0.0031$&$0.0024$&$0.0025$\\
$11$ &\phantom{-}$11.0077$&\phantom{-}$11.0224$&\phantom{-}$11.0224$          &$0.0000$&  $0.0018$&$0.0019$&$0.0019$\\
$12$ &\phantom{-}$12.0073$&\phantom{-}$12.0155$&\phantom{-}$12.0155$          &$0.0000$&  $0.0008$&$0.0014$&$0.0014$\\
$13$ &\phantom{-}$13.0056$&\phantom{-}$13.0093$&\phantom{-}$13.0093$          &$0.0000$&  $0.0005$&$0.0010$&$0.0010$\\
$14$ &\phantom{-}$14.0053$&\phantom{-}$14.0065$&\phantom{-}$14.0066$          &$0.0000$&  $0.0003$&$0.0006$&$0.0006$\\
$15$ &\phantom{-}$15.0043$&\phantom{-}$15.0044$&\phantom{-}$15.0043$          &$0.0000$&  $0.0002$&$0.0003$&$0.0003$\\
\hline
\end{tabular}
\end{center}
\end{table}

\subsection{Experimental phase shifts}

Our  goal is to assess the effective central potential governing
the $p-\alpha$ scattering events. Comprehensive data of phase
shift analysis of the $p-\alpha$ scattering has been presented by
Ali, Ahmad and Ferdous in \cite{Ali}. Because of the spin-orbit
coupling both spin-up $\delta_l^+$  and spin-down $\delta_l^-$
phase shifts contribute to the scattering amplitude at each
partial wave. In case of weak spin-orbit coupling the combined
phase shifts
\begin{equation}
\hat\delta_l=\frac{1}{2l+1}[(l+1)\delta_l^+ +l\delta_l^-]
\end{equation}
are characteristic of  the underlying central potential
\cite{Leeb1995}, and we shall use them as input for the CT
procedures.

We have inverted the  phase shift data of Ali {\it et al}. Since
the inverse potentials exhibit similar characteristics we present
only some  of the inversion results here.

\begin{figure}[ht]
\begin{center}
\rotatebox{0}{
 \includegraphics[width=11cm]{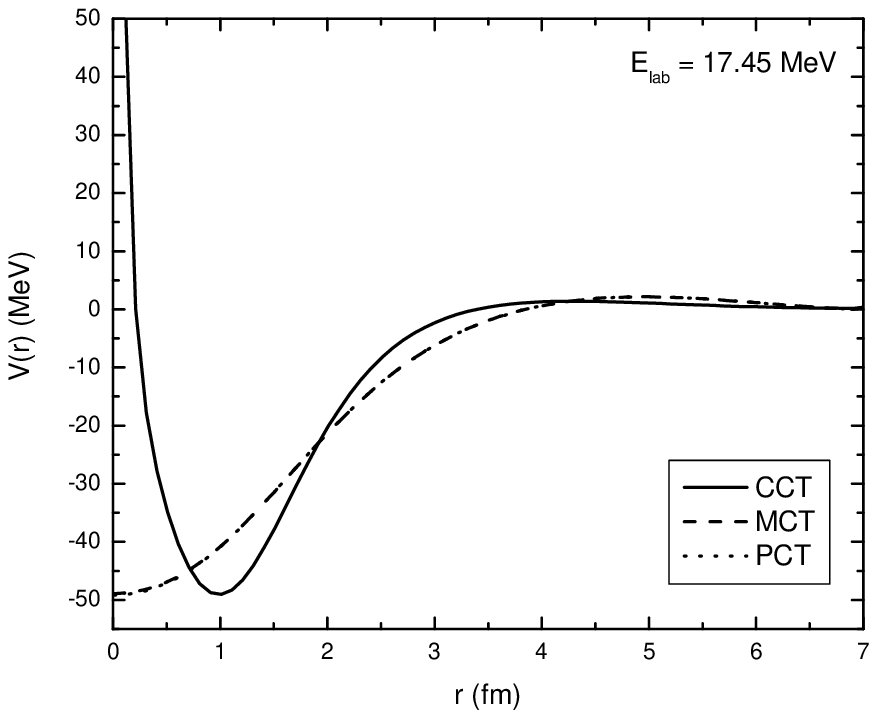}
}
   \caption{Inverse $p-\alpha$ potentials $V(r)$ obtained from input phase shifts
   $\hat{\delta}_l^{\rm{orig}}$ (given in table \ref{tab1}) as a function of the radial distance $r$   at  energy $E_{\rm{lab}}=17.45$ MeV ($E_{\rm{c.m.}}=13.96$ MeV, $k=0.731$ fm$^{-1}$). Curves  obtained  by  the PCT, MCT and CCT method are labeled accordingly.}
\label{fig1}
\end{center}
\end{figure}

In  figure \ref{fig1} the inverse potentials yielded by three CT methods
at $E_{\rm lab}=17.45$ MeV proton energy are depicted. The results
at this energy are representative of the potentials recovered
below the $E_{\rm lab}=22.94$ MeV, $\alpha+p\rightarrow d+^{3}$He
inelastic threshold. As we see the MCT and PCT potentials are
almost identical and strongly resemble a Woods-Saxon form. The
range and strength of all the three potentials are similar but the
CCT potential is different in shape: it possesses a repulsive
core.

Note  that a repulsion at small distances can be theoretically
accounted for as the manifestation of the Coulombic non-locality
\cite{bachelet1982}. It can also be  seen that  this repulsion
core stabilizes the inverse potential in the sense that the
amplitude of the asymptotic oscillations is diminished compared to
the non-repulsive MCT/PCT results. This is also the reason why the
phase shift reproduction with a given precision of the CCT
potential is better than that of the MCT/PCT potentials (see table \ref{tab1}).

\begin{figure}[ht]
  \begin{minipage}[b]{0.5\linewidth}
     \centering
     \subfloat[][Real part of the CCT potential.]{\includegraphics[width=7cm]{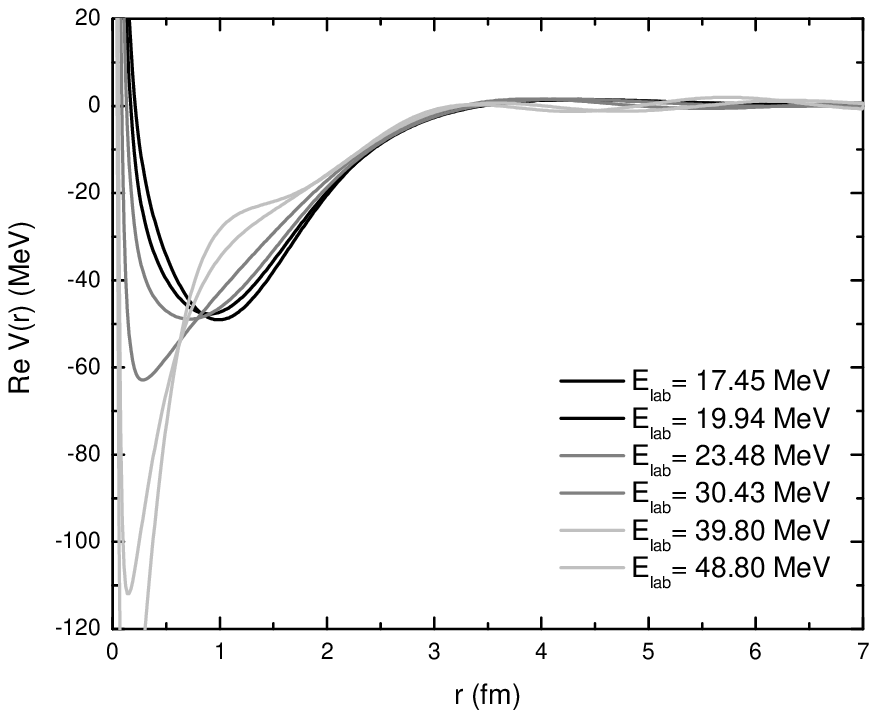}}\\
     \subfloat[][Imaginary part of the CCT potential.]{\includegraphics[width=7cm]{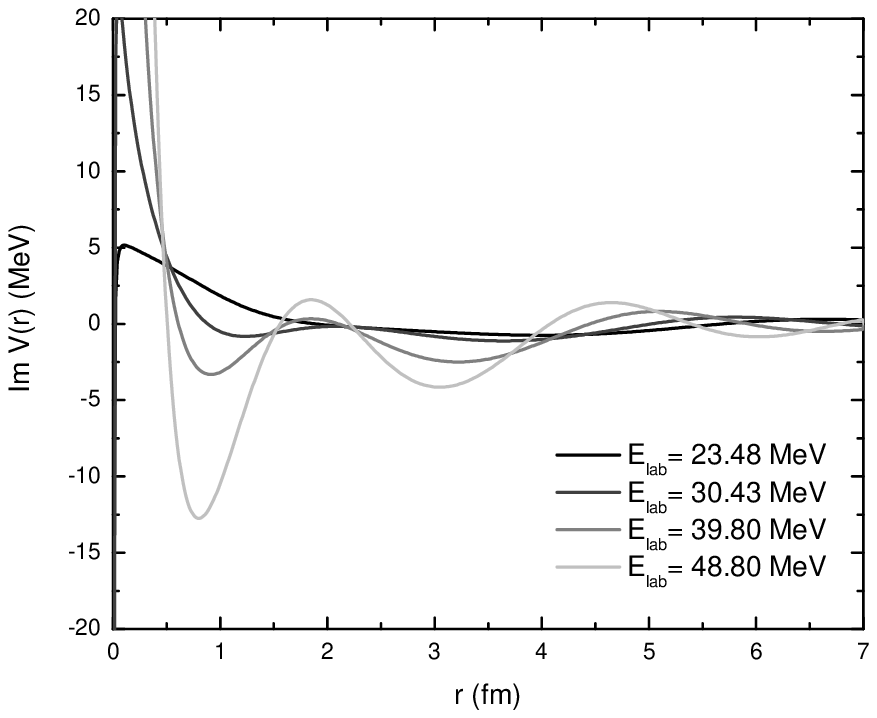}}\\
  \end{minipage}
  \hspace{0.1cm}
  \begin{minipage}[b]{0.5\linewidth}
     \centering
       \subfloat[][Real part of the PCT potential.]{\includegraphics[width=7cm]{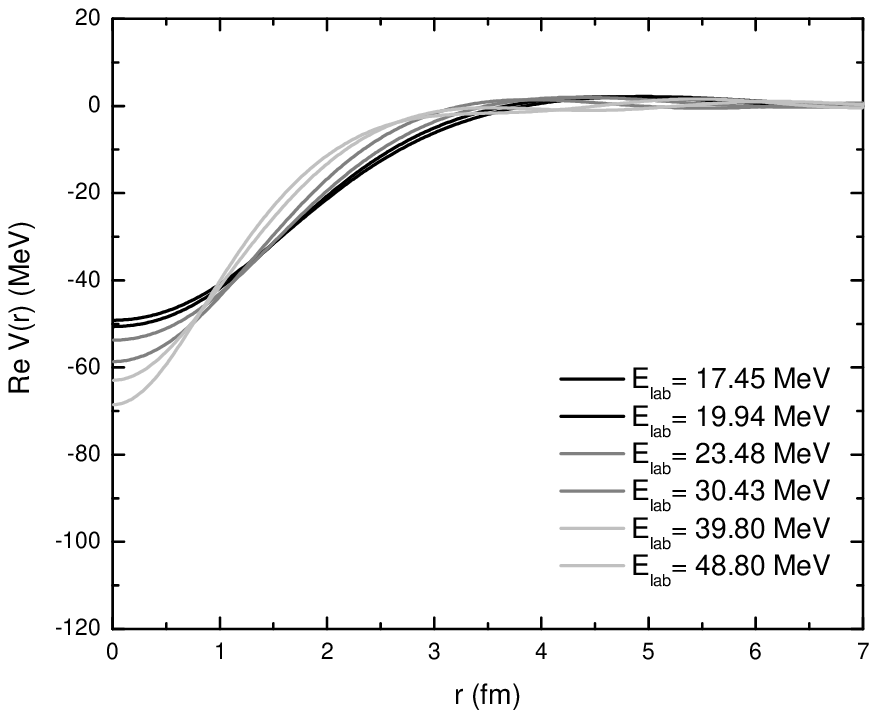}}\\
     
     \subfloat[][Imaginary part of the PCT potential.]{\includegraphics[width=7cm]{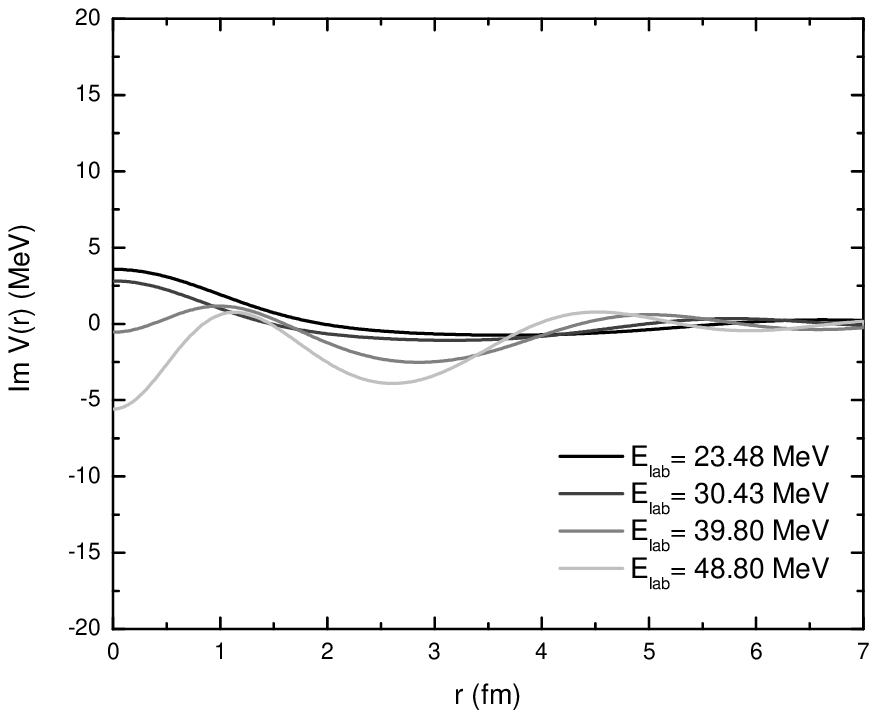}}\\
   \end{minipage}
   \caption{Complex-valued inverse $p-\alpha$ potentials yielded by the CCT (a,b) and PCT (c,d) methods at various $E_{\rm lab}$
   proton energies below and above the inelastic threshold using the experimental phase shifts of \cite{Ali}.} \label{fig2}
\end{figure}

In figure \ref{fig2} numerous other CCT and  PCT potentials are shown at
various energy values including those above inelastic threshold.
(The MCT potentials are not shown because they coincide within the
width of line with the PCT results.) Apart from the Coulombic
singularity at the origin, the potentials have a similar range of
$3-4$ fm and strength of $50-70$ MeV (PCT) and $50-160$ MeV (CCT).
The imaginary part is much less compared to the real part. The
reproduction of phase shifts (not shown)  gets better at higher
energy in case of the PCT potentials because of the fixed cut-off
radius which in principle does not apply to CCT  (and MCT) method.
Without use of this radius the CCT (and MCT) potentials give back
the input phase shifts exactly.

\begin{table}[h]\label{tab1}
\caption{ Inversion results of the experimental
$p-\alpha$ data at $E_{\rm{lab}}=17.45$ MeV. The matching parameter
used in the PCT  and MCT procedures was set to $r_{\rm a}=7$ fm from
which distance also the MCT and CCT potentials have been replaced
by the pure Coulombic tail.}

\begin{center}
\begin{tabular}{cccccccc}

\hline
$l$ & $L^{{\rm CCT}}$ & $L^{{\rm MCT}}$ & $L^{{\rm PCT}}$ & $\hat{\delta}_l^{{\rm orig}}$ & $\Delta\hat{\delta}_l^{{\rm CCT}}$&$\Delta\hat{\delta}_l^{{\rm MCT}}$&$\Delta\hat{\delta}_l^{{\rm PCT}}$\\
\hline
$0$  &$-1.7204$&$-1.6703$&$-1.6705$          &$1.7240$&$0.0139$&$0.0219$&$0.0198$\\
$1$ & $\phantom{-}$$0.5252$& $\phantom{-}$$0.4537$ &$\phantom{-}$$0.4534$           &$1.4839$&$0.0249$&$0.0589$&$0.0597$\\
$2$ & $\phantom{-}$$2.0328$& $\phantom{-}$$2.0210$ &$\phantom{-}$$2.0211$           &$0.0760$&$0.0206$&$0.0107$&$0.0108$\\
$3$  &$\phantom{-}$$3.0475$& $\phantom{-}$$3.0881$ &$\phantom{-}$$3.0882$           &$0.0229$&$0.0250$&$0.0762$&$0.0765$\\
\hline
\end{tabular}
\end{center}
\end{table}

\section{Conclusions, summary}

We have developed  the Cox-Thompson (CT) fixed energy inverse
quantum scattering method into various directions in order to make
it appropriate for treating long range interactions. By explicit
calculation we have shown that the modified (MCT) and phase
transformed (PCT) methods yield practically the same potentials
although quite other functions are involved in the calculation.
These potentials are generally finite at the origin therefore they
can be used for cases with no singularity at small distances. The
Coulomb (CCT) method produces a characteristic Coulomb singularity
at the origin. This method can be applied with success when
nonlocality or a repulsion hard core plays a role in the
interaction.

We have applied  the methods first to model cases with and without
singularity. The model potentials describing $\alpha-\alpha$
scattering are well reproduced by the new methods according to
their characteristic properties concerning the finiteness or
infiniteness at the origin. Then, the experimental $p-\alpha$
phases of Ali {\it et al} \cite{Ali} have been inverted with the
result that the CCT potentials showing up a repulsive hard core
reproduce better the input phase shifts than the PCT potentials
which exhibit a Woods-Saxon shape. Whether or not the repulsive
core is present between the proton and alpha particle when being
scattered by each other should be clarified also using independent
source of information (see e.g. \cite{Ali,Gammel,Satchler,Thompson1,Thompson2}).

\section*{Acknowledgements} 
The authors thank the DFG for supports through the contract No. \mbox{436 UNG 113/201/0-1}.

\appendix

\section{Coulomb functions of complex order}

In certain applications of both the CCT and MCT formulations of the Cox-Thompson inverse scattering method for long-ranged potentials it is necessary to evaluate the regular and irregular Coulomb wave functions for complex orders. Based on \cite{coulFG,coulFG2} the evaluation is accomplished by using the power series given below.

The regular Coulomb wave function for $L\in\mathbb{C}$ complex order is given by
\begin{equation}
 F_L(x)=\frac{2^L d^{-\frac{\pi\eta}{2}}\left[\Gamma(L+1+ i\eta)\Gamma(L+1- i\eta)\right]^{\frac{1}{2}}}{\Gamma(2L+2)}\cdot x^{L+1}\sum_{j=L+1}^\infty C_j^L(\eta)x^{j-L-1},
\end{equation}
and the $C_j^L$ constants are defined by the recursion
\begin{eqnarray}
C_{L+1}^L&=1,\\
C_{L+2}^L&=\frac{\eta}{L+1},\\
C_j^L&=\frac{2\eta C_{j-1}^L-C_{j-2}^L}{(j+L)(j-L-1)}\qquad j>L+2.
\end{eqnarray}
This production of the $F_L(x)$ is a simple analytic continuation of the formulae in \cite{abramo}.

However the irregular Coulomb wave function cannot be given by such a simple generalization. We utilize therefore the fact that the Coulomb wave functions and the Whittaker functions are related to each other linearly for fixed $L$ and $\eta$. By means of the Whittaker functions the regular Coulomb wave function is also given.
\begin{eqnarray}
F_L(x)&=A(L,\eta)M\left( i\eta,L+\frac{1}{2},2 i x\right),\\
G_L(x)&= i A(L,\eta)M\left( i\eta,L+\frac{1}{2},2 i x\right)+B(L,\eta)W\left( i\eta,L+\frac{1}{2},2 i x\right)
\end{eqnarray}
with
\begin{eqnarray}
&A(L,\eta)=\frac{ d^{- i(L+1)\frac{\pi}{2}} d^{-\frac{\pi\eta}{2}}[\Gamma(L+1+ i\eta)\Gamma(L+1- i\eta)]^{\frac{1}{2}}}{2\Gamma(2L+2)}\\
&B(L,\eta)=\frac{ d^{+ i L\frac{\pi}{2}} d^{+\frac{\pi\eta}{2}}[\Gamma(L+1+ i\eta)\Gamma(L+1- i\eta)]^{\frac{1}{2}}}{\Gamma(L+1+ i\eta)}.
\end{eqnarray}
The Whittaker functions $M(a,b,y)$ and $W(a,b,y)$ are well-known for complex arguments (see \cite{abramo} chapter 13). The Gamma function was calculated by a Lanczos series approximation.

The $\eta$ parameter appearing in the above formulae is the Sommerfeld parameter given with the  quantities  discussed earlier as
\begin{equation}
\eta=\frac{k}{E}\frac{e^2}{8\pi\varepsilon_0}Z_1Z_2=\frac{\sqrt{2}}{\hbar}\frac{e^2}{8\pi\varepsilon_0}\cdot\sqrt{\frac{m}{E}}Z_1Z_2.
\end{equation}

\end{document}